\documentclass[iop]{emulateapj-rtx4}
\usepackage{graphicx,amsmath,amssymb}
\usepackage{natbib}
\usepackage{apjfonts}
\usepackage{multirow}

\usepackage{booktabs}

\begin{document}

   \title{Investigating CXOU~J163802.6--471358: A new pulsar wind nebula in the Norma region?}

   \author{Simone J. Jakobsen$^1$}\author{John A. Tomsick$^2$}\author{Darach Watson$^1$}\author{Eric V. Gotthelf$^3$}\author{Victoria M. Kaspi$^4$}\affil{$^1$Dark Cosmology Centre, Niels Bohr Institute, University of Copenhagen, Juliane Maries Vej 30, DK-2100 Copenhagen \O, Denmark;\\\texttt{}} 
   \affil{$^2$Space Sciences Laboratory, 7 Gauss Way, University of California, Berkeley, CA 94720-7450, USA ;\\\texttt{}} \affil{$^3$Columbia Astrophysics Laboratory, Columbia University, New York, NY 10027, USA;\\\texttt{}} \affil{$^4$Department of Physics, McGill University, Montreal, Quebec H3A 2T8, Canada;\\\texttt{}}

   \begin{abstract}
We present the first analysis of the extended source CXOU~J163802.6--471358, which was discovered serendipitously during the {\em Chandra} X-ray survey of the Norma region of the Galactic spiral arms. The X-ray source exhibits a cometary appearance with a point source and an extended tail region. The complete source spectrum is fitted well with an absorbed power law model and jointly fitting the {\em Chandra} spectrum of the full source with one obtained from an archived {\em XMM-Newton} observation results in best fit parameters $N_{\rm H}$ $=1.5^{+0.7}_{-0.5}\times10^{23} \text{cm}{^{-2}}$ and $\Gamma=1.1^{+0.7}_{-0.6}$ (90$\%$ confidence uncertainties). The unabsorbed luminosity of the full source is then $L_X\sim 4.8\times10^{33}d_{10}^2$ergs s$^{-1}$ with $d_{10}=d/10$kpc, where a distance of 10 kpc is a lower bound inferred from the large column density. The radio counterpart found for the source using data from the Molonglo Galactic Plane Survey epoch-2 (MGPS-2) shows an elongated tail offset from the X-ray emission. No infrared counterpart was found. The results are consistent with the source being a previously unknown pulsar driving a bow shock through the ambient medium.
   \end{abstract}
   \keywords{acceleration of particles --- X-rays: general --- radio continuum: general --- pulsars: general --- stars: individual(CXOU J163802.6-471358)
             }

   \maketitle

%
%
\section{Introduction\label{introduction}}
Most pulsars are discovered in radio surveys by their pulsation properties. When they are later resolved in X-rays with high resolution telescopes such as the {\em Chandra X-ray Observatory}, complex structures associated with the pulsar are discovered. Since the pulsar loses rotational energy through a wind of ultrarelativistic particles, the non-thermal synchrotron emission that arises will be observable across the electromagnetic spectrum in the form of a pulsar wind nebula (PWN; see \citet{gaensler2006} and \citet{kirk2009} for a review). All pulsars are expected to be associated with a PWN \citep{kargaltsev2008a}, but only a fraction of them are detected at current sensitivity. The PWNe that have been identified exhibit a morphology dependent on the ambient medium, and on pulsar properties such as age, speed, and magnetic field strength. The high sensitivity and arcsecond angular resolution of the {\em Chandra} telescope makes it possible to study the X-ray morphology of these sometimes weak extended sources as well as their spectral properties. For a pulsar moving supersonically through the ambient medium, the PWN will be confined by ram pressure and display a cometary structure in X-rays. This type is generally referred to as a bowshock-tail PWN, where "The Mouse" PWN \citep{gaensler2004} is a typical example. \\
\indent An X-ray survey of a 2$^\circ\times0.8^\circ$ region of the Norma spiral arm was done during summer 2011. Twenty-seven observations, each of $\sim$ 20 ks duration, were performed with the {\em Chandra X-ray Observatory} for an in depth study of the region. The primary goal of the survey was to find new High-Mass X-ray Binaries (HMXBs) in order to study their evolution. The Norma region was chosen because the hard X-ray imaging of the Galactic plane done by the {\em INTERnational Gamma-Ray Astrophysics Laboratory (INTEGRAL)} revealed that the Norma region contains the largest number of HMXBs and OB associations after the Galactic center (Fornasini et al., submitted). Around 1500 sources were found in the {\em Chandra} survey and a catalog of these sources is presented in another paper (Fornasini et al., submitted). Out of these 1500 sources at least 6 are extended. Four of these have been previously detected and categorized as Supernova Remnants (SNRs) and are mentioned in SNR catalogs such as the Green catalog \citep{green2009} and the new high energy Galactic SNR catalog \citep{ferrand2012}. One of the other two new sources is treated in this paper. \\
\indent In Section \ref{observations} we present the data that was utilized in this investigation. In Sections \ref{image} and \ref{spec} we present the methods used and the results that were achieved based on the X-ray data available. Section \ref{multi} describes the attempt to identify counterparts to the X-ray emission using infrared and radio data. Finally, in Section \ref{discussion} we evaluate the results in order to identify the source type.

%
%

\section{Observations}\label{observations} 
The extended source CXOU~J163802.6--471358 was serendipitously covered by two {\em Chandra} exposures with ObsIDs 12519 and 12520 (19.3 ks and 19.0 ks exposures, respectively). Both were performed on 2011 June 13 using the ACIS-I instrument in VFAINT mode, and both had the source positioned off-axis, more so in ObsID 12520, where it is located at $\theta \sim 8.4\arcmin$ compared to $\theta \sim 3.8\arcmin$ for ObsID 12519. \\
\indent All {\em Chandra} data preparation was performed using the {\em Chandra} Interactive Analysis of Observations (CIAO) package v4.4. The event files to be used for imaging analysis were filtered on energy so as to only include energies in the range of 0.5$-$8.0 keV. In order to increase the signal to noise ratio, the two observations were merged using the CIAO routine {\ttfamily merge$\_$obs}, providing a merged exposure corrected image of the source. The spectral analysis, on the other hand, was done using the unfiltered level 2 event files. \\
\indent Contaminating point sources of significance larger than 4$\sigma$ as listed in the Norma region point source catalog (Fornasini et al., submitted) were removed prior to the analysis.\\
\indent CXOU~J163802.6--471358 was found to be covered by an archived {\em XMM-Newton} observation from 2005 August 19-20, and the source has been cataloged as 3XMM~J163802.6--471357. In this work, we include the data from this observation (ObsID 0307170201), which consists of a 99.5\,ks pointing during which CXOU~J163802.6--471358 was 11.5$^{\prime}$ off-axis.  Including the {\em XMM-Newton} data allows us to compare to previous measurements and to improve the constraints on the spectral parameters. \\
\indent The source field has also been covered by several surveys over multiple wavelengths, as both radio data from the Molonglo Galactic Plane Survey epoch 2 (MGPS-2) \citep{murphy2007} and infrared observations from the {\em Spitzer} space telescope \footnote{http://irsa.ipac.caltech.edu/data/SPITZER/GLIMPSE/} and the Visible and Infrared Survey Telescope for Astronomy (VISTA)\footnote{http://horus.roe.ac.uk/vsa/index.html} are available. We make use of these survey images in order to look for counterparts to the X-ray source.\\

\section{Analysis and Results}\label{analysis}
\begin{figure}[tb]
\centering
\includegraphics[trim=50 10 93 50,clip,width=\columnwidth]{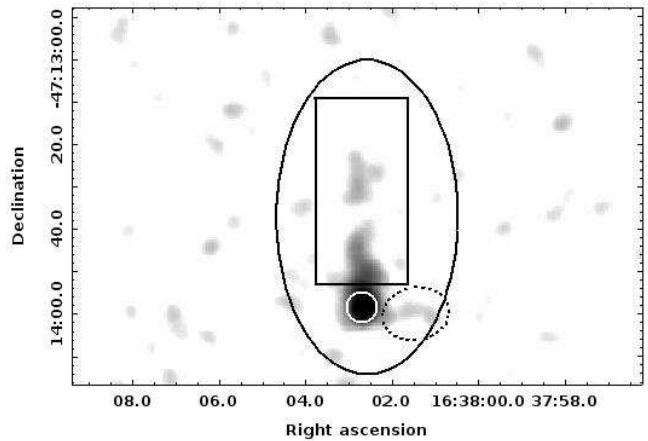}
\caption{The image shown is an exposure corrected merged image of the PWN candidate. The solid ellipse defines the full source area, the white circular region defines the point source portion of the source and the rectangular region defines the tail section. The diffuse feature (marked with a dashed ellipse) has a significance just below the detection threshold. The solid line regions were used for spectral extraction. North is up and East is to the left.}
\label{fig:specreg}
\end{figure}

A quick examination of the merged {\em Chandra} image (Figure \ref{fig:specreg}) reveals a complex structure of CXOU~J163802.6--471358 featuring two distinct components: a point-like part of detection significance 38.9$\sigma$; and a diffuse part with significance 9.9$\sigma$, both calculated for the energy range 2.0$-$8.0 keV. This energy range was chosen since no source counts were detected below 2.0 keV, presumably as a result of the high column density towards the Norma region. The point-like source is positioned at $(\alpha,\delta)_{\text{J2000}} = (16^{\text{h}}38^{\text{m}}02^{\text{s}}.7,-47^{\circ}13'58''.4)$ in equatorial coordinates with a position error of $0.6\arcsec$ (at $95\%$ confidence) (Fornasini et al., submitted) and the source significance was calculated based on the events falling inside a circular region centered on that position, having a radius of 3.5$\arcsec$ (the white circular region in Figure \ref{fig:specreg}). The significance of the diffuse emission was calculated from the events falling inside the rectangular region shown in Figure \ref{fig:specreg}. Since the significance is calculated based on the summed counts in both exposures, the emission from the point-like source is somewhat blended with the diffuse extended emission, and we chose the rectangular region so that it included as many counts as possible without including too many of the point source photons. The diffuse part is tail-like (referred to as "the tail" hereafter) and extends from the point-like source $\approx40\arcsec$ to the north. \\
\indent Furthermore, when smoothing the image, two diffuse features emerge. One of them is only visible in ObsID 12520 (marked with a dashed ellipse in Figure \ref{fig:specreg}), and its significance is right below the limit for positive detection at 2.9$\sigma$, and stretches $\approx19.5\arcsec$ to the West. The other one is possibly observed in ObsID 12519 below the significance threshold at 2.4$\sigma$, but is not visible in the merged image (Figure \ref{fig:specreg}), since it is blended with the PSF of observation 12520. \\

\subsection{Imaging analysis}\label{image}
\begin{figure}[tb]
\centering
\includegraphics[trim=0 0 0 0,clip,width=\columnwidth]{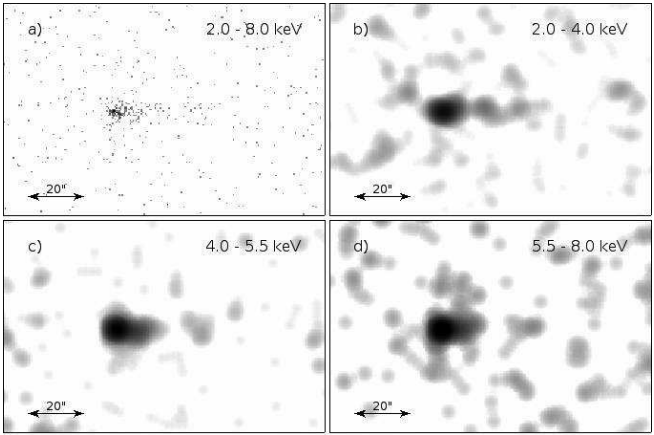}
\caption{The PWN candidate is shown here in four different energy ranges. All four images are exposure corrected and unbinned with pixelsize $\sim 0.492''\times0.492''$. Panel b)-d) have additionally been smoothed using a Gaussian kernel of size $\sigma=2.5\arcsec$. North is to the right and East is up.}
\label{fig:bands}
\end{figure}

\begin{figure}
\centering
\includegraphics[trim=30 20 50 20,clip,width=\columnwidth]{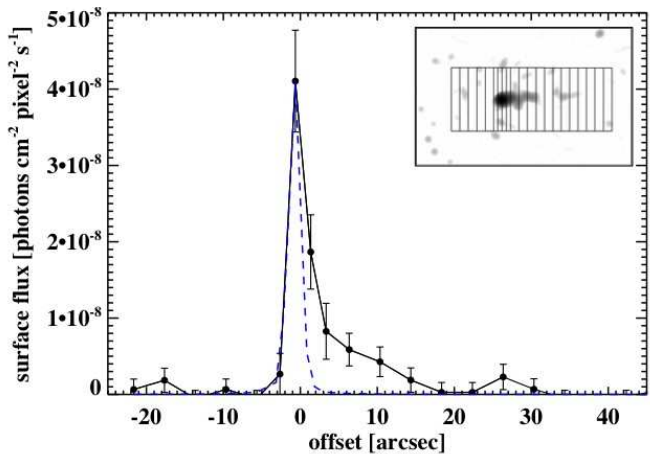}
\caption{The figure shows the background subtracted and exposure corrected surface brightness profile of CXOU~J163802.6--471358 created using aperture slices along the symmetry axis. The profile only includes events from obsID 12519 to avoid blending issues. The energy range is 2.0$-$8.0 keV. Plotted on top (dashed, blue line) is the {\em Chandra} PSF simulated at the same off-axis angle as that of the observation. An inset image shows the extraction regions used.}
\label{fig:slices}
\end{figure}

An exposure corrected, full resolution (0.492$\arcsec\times$0.492$\arcsec$), merged {\em Chandra} image of CXOU~J163802.6--471358 for the energy range 2.0-8.0 keV is shown in the top left panel of Figure \ref{fig:bands}.
The source is hard, peaking at an energy of $\sim$ 5 keV, and in order to examine the morphology of the source with energy, a merged, exposure corrected image was additionally created for three narrow energy bands, representing soft, medium and hard X-ray energies. The energy bands were chosen so that a similar number of source net counts ($\sim$100) were present in each band. Since no source counts were detected below 2.0 keV, the selected energy ranges were 2.0-4.0 keV (\emph{soft}), 4.0-5.5 keV (\emph{medium}) and 5.5-8.0 keV (\emph{hard}), as shown in Figure \ref{fig:bands}.\\
\indent The spatial extension of the source was examined further through a surface brightness profile constructed using aperture slices taken along the symmetry axis. Only observation 12519 was used in this step, since the tail and point-like source were expected to be somewhat blended in observation 12520, caused by the location of the source at a large off-axis angle. The resulting profile can be seen in Figure \ref{fig:slices} with the inset image showing the extraction regions used. Each rectangular region measures $30\arcsec\times4\arcsec$ across the tail section and to the South of the point-like source, whereas the regions lying on top of the point-like source measures $30\arcsec\times2\arcsec$. \\
\indent In order to determine whether the main component is in fact a point source, a {\em Chandra} PSF was simulated at the source position, and superposed on the profile. The PSF was generated using the Chandra Ray-Tracing (ChaRT) software\footnote{http://cxc.harvard.edu/chart/runchart.html} and simulated using the MARX simulator v5.0.\\

\subsection{Spectral analysis}\label{spec}
\begin{figure}[tb]
\centering
\includegraphics[trim=0 6.4 75 0,clip,width=\columnwidth]{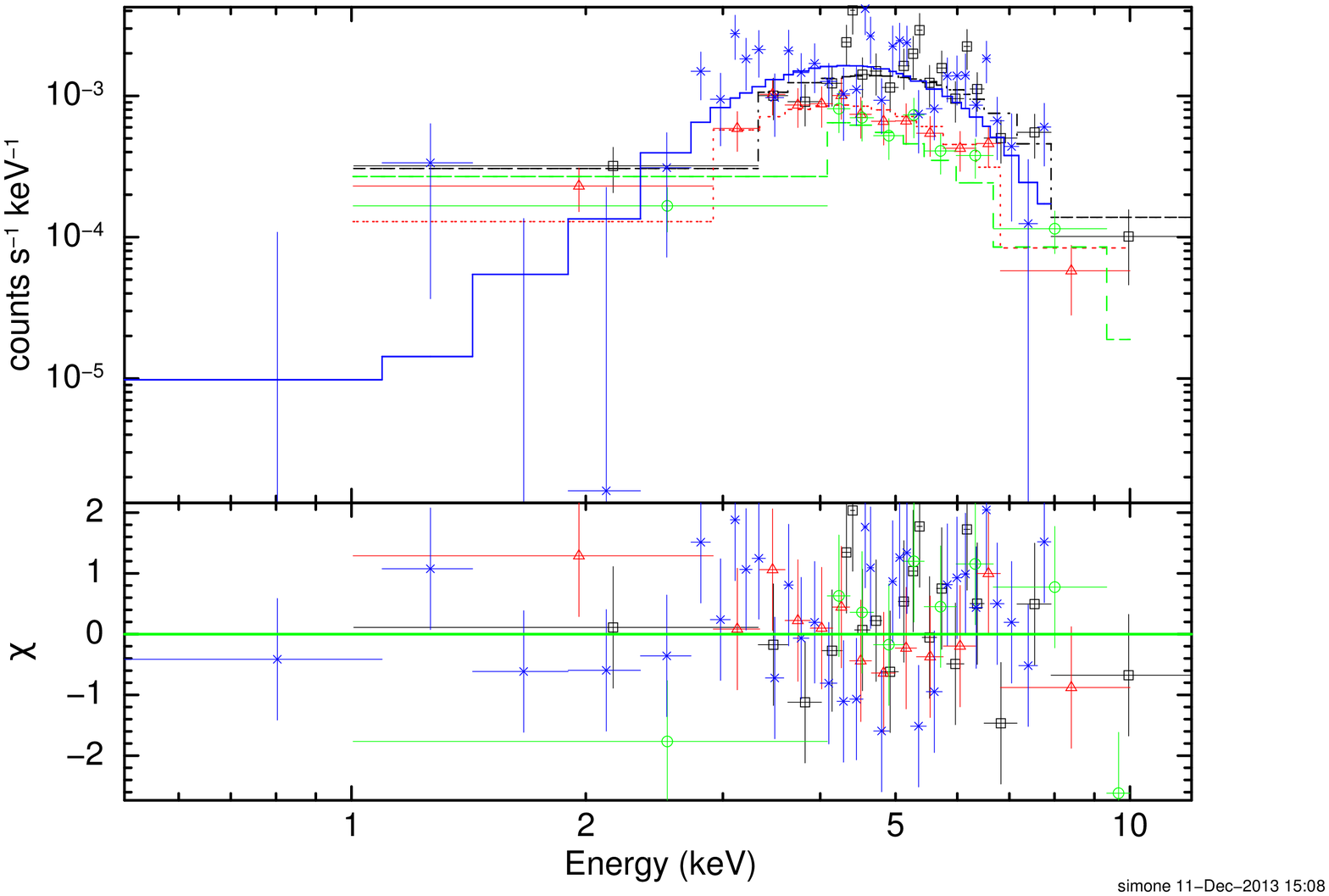}
\caption{The simultaneous fit of the three {\em XMM} spectra obtained from the PN (squares, dash-dotted black line), MOS1 (triangles, dotted red line), MOS2 (circles, dashed green line) instrument, and of the {\em Chandra} spectrum (stars, solid blue line). The spectra were fitted in XSPEC using an absorbed power-law model. Color figure available in online version.}
\label{fig:spec}
\end{figure}

In order to look for spectral variations across the {\em Chandra} source, we extracted spectra from the two source components (the point source and the tail) along with a spectrum of the full source area. The extraction regions were defined as shown in Figure \ref{fig:specreg}, and all spectra and the associated RMF and ARF files were created using the CIAO tool {\ttfamily specextract}. Events from both observations 12519 and 12520 were extracted for each region, and combined to remediate the sparse number of counts in each exposure. One background spectrum was also extracted for each exposure from a source-vacant area on the chip with the source, and combined using the procedure described above. All channels below 0.5 keV and above 8.0 keV were ignored, and using the {\ttfamily grppha} tool from the HEASARC package, the remaining source counts were subsequently grouped, so that each bin contained a minimum of 10 counts.\\
\indent The three resulting spectra were loaded into XSPEC and fitted simultaneously using an absorbed power-law model with the $N_{\rm H}$ parameter for the three tied together and the photon index varying freely. This was done assuming that the column density $N_{\rm H}$ does not vary across the source, which is in good agreement with the fact that the best fit value of the $N_{\rm H}$ parameter from individual fits is the same within 90\% confidence. The resulting best fit parameters for the three regions can be seen in the top half of Table \ref{tab:spec}. \\
\indent After fitting the grouped spectra, we examined whether using Cash statistics \citep{cash1979} on the ungrouped spectra gave a significantly different result. The fit parameters were consistent within 90\% confidence and the results based on the grouped spectra were therefore used in the further analysis.\\
\indent Given the generally poor photon statistics, a hardness ratio was additionally calculated for each source component using the general formula $HR = (H-S)/(H+S)$, where $H$ is the net count number in the hard energy band $4.5-8.0$ keV and $S$ is the net count number in the soft energy band $2.0-4.5$ keV. The energy ranges were again chosen so that approximately the same number of source counts from the full source field reside in each band. The calculation resulted in a hardness ratio with 1$\sigma$ errors of $0.16\pm0.10$ and $0.08\pm0.21$ for the point source and the tail component, respectively, meaning that there is no significant difference between the hardness of the two.\\
\indent We made an {\em XMM-Newton} spectrum for the PWN candidate using the Scientific Analysis Software (SAS) v13.0.1. We produced new event files using {\ttfamily epproc} and {\ttfamily emproc} and filtered them according to the recommended procedures.\footnote{See http://xmm.esac.esa.int/sas/current/documentation/threads/.}  Based on the 10--12\,keV PN light curve, we removed events that occurred during background flares. Such flares were prevalent during this observation, and we were only able to use approximately half of the time, leaving exposure times of 48.5\,ks for PN and 54.2\,ks for each MOS unit.  The angular resolution of {\em XMM-Newton} does not allow us to separate the point source from the extended emission, and, for all three instruments (PN, MOS1, and MOS2), we extracted a spectrum from a $30^{\prime\prime}$-radius circle, which includes both the point source and the tail.  In order to subtract the background, we used nearby rectangular regions that do not include sources.  We re-binned the spectra to require a signal-to-noise ratio of at least 2.7 in each bin (except for the highest energy bin).  We obtained source count rates of $6.3\times 10^{-3}$, $3.2\times 10^{-3}$, and $2.2\times 10^{-3}$ counts\,s$^{-1}$ for PN, MOS1, and MOS2, respectively.\\
\indent The three spectra from the three instruments (PN, MOS1 and MOS2) were subsequently loaded into XSPEC and fitted simultaneously using the same absorbed power-law model as used for the {\em Chandra} ACIS spectra. The column density $N_{\rm H}$ was tied together and all parameters including an inter-instrumental normalization constant were allowed to vary freely. The best fit parameters resulting from the fit can be seen in the bottom part of Table \ref{tab:spec}. \\
\indent The {\em Chandra} and {\em XMM-Newton} best fit parameters were found consistent within 90$\%$ confidence, and based on this, the {\em Chandra} full source spectrum and the three {\em XMM-Newton} spectra were fitted simultaneously following the same fitting procedure as previously. The fitted full source spectra and their residuals can be seen in Figure \ref{fig:spec}, and the best fit parameters are listed in Table \ref{tab:spec}.\\

\begin{table*}[!ht]
\centering
\caption{Spectral fitting results for CXOU~J163802.6--471358}
\begin{minipage}{11cm}
\begin{tabular}{lcccccr}
\toprule\toprule
Region    & $N_{\rm H}$$^a$    &  $\Gamma$    &  Goodness of fit  &  $F^{obs}_{(2-10\text{ keV})}$    &  counts \\
       &    [$10^{22}$cm$^{-2}$]      &           &       [$\chi^2$/d.o.f.]        &     [10$^{-13}$ergs/cm$^2$/s]       &      \#     \\
\midrule
1 - full source & $13^{+6}_{-5}$ & $0.7^{+1.0}_{-0.9}$ &  60.5/54  &  3.2$^{+0.9}_{-1.0}$ &  356  \\
\noalign{\smallskip}
2 - point source  &  ... &  $0.5^{+1.0}_{-0.9}$ &   ...  &  2.5$^{+0.7}_{-0.8}$   &  180  \\
\noalign{\smallskip}
3 - tail  &  ...   &   $1.3^{+1.4}_{-1.2}$    &  ...    &     0.8$^{+0.4}_{-0.4}$   &  110  \\
\noalign{\smallskip}
\midrule\midrule
{\em XMM} full source   &     $19^{+11}_{-8}$   &    $1.4^{+0.9}_{-0.8}$    &     38.4/37      &       2.6$^{+0.3}_{-0.8}$     &     306$^b$  \\
\noalign{\smallskip}
{\em XMM}+{\em Chandra}	&	15$^{+7}_{-5}$  	&	1.1$^{+0.7}_{-0.6}$	&	77.8/70	&	2.6$^{+0.4}_{-0.5}$$^c$	&	...	\\	
\bottomrule
\end{tabular}
\textsc{Note} - Uncertainties are at the 90$\%$ confidence level.\\
$^a$Based on the abundance values given in \citet{wilms2000}.\\
$^b$Number of counts quoted is for the PN spectrum.\\
$^c$The absorbed flux value quoted here is calculated based on the PN spectrum. The value calculated from the ACIS spectrum and the MOS2 spectrum are consistent with this value, whereas there is a slight discrepancy between them and the MOS1 value.
\label{tab:spec}
\end{minipage}
\end{table*}

\subsection{Radio and Infrared Imaging}\label{multi}
\begin{figure}[tb]
\centering
\includegraphics[trim=10 2 40 30,clip,width=\columnwidth]{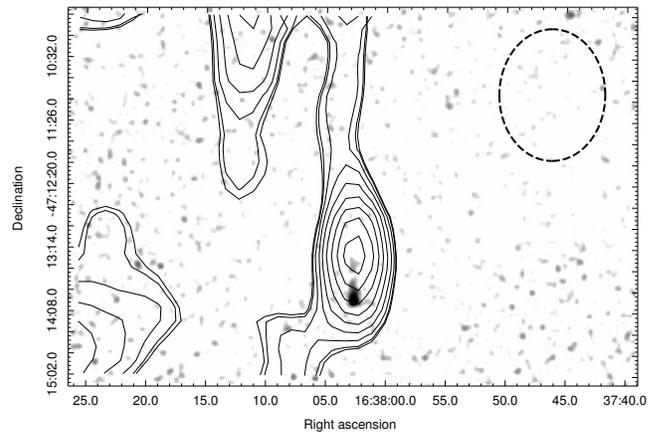}
\caption{Shown here is the exposure corrected merged image of CXOU~J163802.6--471358 in greyscale, smoothed with a Gaussian kernel of size $\sigma$ = 2$\arcsec$. On top is an overlay of radio contours from the MGPS-2 showing 7-70 mJy beam$^{-1}$ in 10 square root steps. The beam shape and size of $45\times45 \hspace{0.1cm}\text{csc} \lvert \delta \rvert \hspace{0.1cm} \text{arcsec}^2$ is shown in the top right corner of the image.}
\label{fig:radio}
\end{figure}

\begin{figure}[tb]
\centering
\includegraphics[width=\columnwidth]{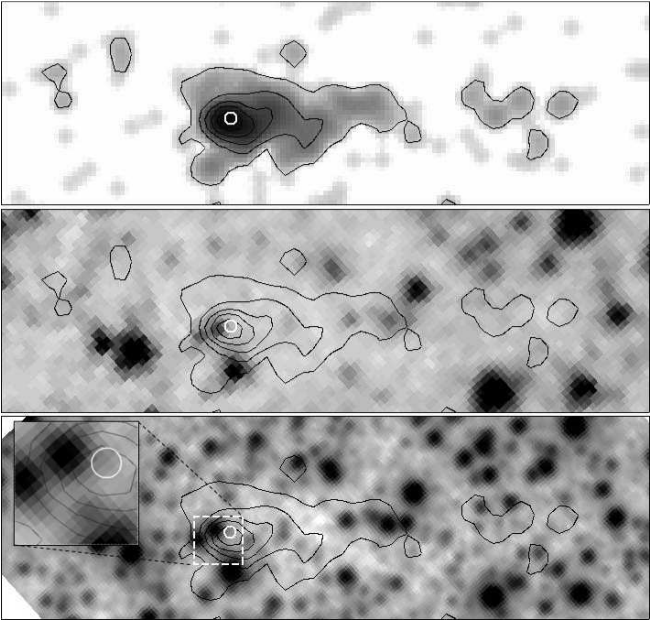}
\caption{An infrared view of the source field. Top plot: {\em Chandra} image with {\em Chandra} contours. The panel measures $\approx1.0\arcmin\times0.4\arcmin$. Middle plot: Infrared 4.5 $\mu$m obtained from the GLIMPSE survey. Bottom panel: Infrared 2.15$\mu$m from VISTA. The circle indicates the position of the {\em Chandra} point source and its 0.6$\arcsec$ position error, which includes the absolute astrometric accuracy. The nearest infrared source is located 1.7$\arcsec$ away in the VISTA survey. The square cut-out in the bottom panel measures 5$\arcsec\times5\arcsec$.}
\label{fig:IR}
\end{figure}

A large part of the Southern Galactic Plane is covered by the Molonglo Galactic Plane Survey epoch 2 (MGPS-2), which includes the Norma region and source CXOU~J163802.6--471358. This provides us with the means to search for a radio counterpart to the X-ray source.\\
\indent The MGPS-2 was done using the Molonglo Observatory Synthesis Telescope (MOST) at a frequency of 843 MHz and a restoring beam size of $45 \times 45 \hspace{0.1cm}\text{csc} \lvert \delta \rvert \hspace{0.1cm} \text{arcsec}^2$, where $\delta$ is the declination. \\
\indent The particular MGPS-2 mosaic image covering the source region was examined, and a radio structure was found coinciding with the X-ray peak position within the resolution of the observation. An image of the X-ray emission with an overlay of radio contours can be seen in Figure \ref{fig:radio}, with the beam shape, and size indicated. The radio structure seems to be an unresolved point source with a tail stretching approximately 3.3$\arcmin$ from the radio peak towards the North. An estimate of the peak brightness of the radio structure is found as the value of the brightest pixel at 65.5 mJy beam$^{-1}$.\\
\indent We furthermore found through archival searches that the source area has been covered by both the {\em Spitzer} Space Telescope and the Visible and Infrared Survey Telescope for Astronomy (VISTA), providing images in the mid- and near infrared. The {\em Spitzer} survey, called the Galactic Legacy Infrared Mid-Plane Survey Extraordinaire (GLIMPSE) \citep{benjamin2003}, covers a large part of the Galactic disk, including the source CXOU~J163802.6--471358. The {\em Spitzer} IRAC instrument collected images in four bands (3.6, 4.5, 5.8, and 8.0 $\mu$m) and a 4.5$\mu$m image of the source region with an overlay of {\em Chandra} X-ray contours is shown as the middle panel of Figure \ref{fig:IR}.\\
\indent The VISTA images were obtained from the VISTA Science Archive using the VISTA Variables in the Via Lactea (VVV) database \citep{saito2012}. The four filters {\em ZYJHK$_\text{S}$}, corresponding to 0.88, 1.02, 1.25, 1.65, 2.15 $\mu$m respectively, were all examined and the $K_\text{S}$ band is shown as the bottom panel of Figure \ref{fig:IR}. The limiting magnitude for the {\em K}$_\text{S}$ band is $\sim 18$~mag \citep{minniti2010}.

%
%

\section{Discussion}\label{discussion}
The source CXOU~J163802.6--471358 is visibly extended across all the energy ranges presented in Figure \ref{fig:bands}. A comparison between the radial profile of the source and a simulated {\em Chandra} PSF (Figure \ref{fig:slices}) confirms that the source consists of two components, a point source and an extended tail. Both components are detected to a high significance of 38.9$\sigma$ and 9.9$\sigma$, respectively.  \\
\indent Based on the extension of the source, a few possibilities were considered for the source type, an AGN with a jet, a LMXB with a jet, and a bow-shock tail PWN. The results presented in the previous section make us prefer the latter for the following reasons.\\
\indent The full source spectrum (point source + tail) is fitted well with an absorbed power-law, pointing towards a synchrotron origin of the emission. The column density at $N_{\rm H} = 1.5\times10^{23}$cm$^{-2}$, resulting from the joint {\em Chandra} and {\em XMM-Newton} fit, is rather high and indicates that the source is located in the far Norma arm. Comparing to another source in the same area, PWN HESS J1640-465 \citep{Lemiere2009}, it is apparent that the Norma arm is a region of high absorption since this source exhibits an X-ray spectrum of comparable column density ($N_{\rm H} = 1.4\times10^{23}$cm$^{-2}$). This is to be expected, since the Norma region is known to host a large number of molecular clouds \citep{dame2001}.\\
\indent The dust extinction in the line of sight towards the Norma region was found through IRSA\footnote{http://irsa.ipac.caltech.edu/applications/DUST/} to be $A_V = 64$ mag \citep{schlafly2011}. Using this together with the relation $N_{\rm H}/A_V = 2.2^{+0.4}_{-0.3} \times 10^{21}$ cm$^{-2}$ mag$^{-1}$ \citep{watson2011} results in an expected column density of $1.4\times10^{23}$ cm$^{-2}$ for the particular region of this source. This makes the column density obtained from fitting a reasonable value for a source in the far Norma arm, but it cannot be excluded that the source could be extragalactic.\\
\indent The distance to the HESS source was estimated based on measurements of H I absorption features towards the system, by comparing the derived velocity to the Galactic rotation curve \citep{Lemiere2009}. The comparable column density of the two sources, implies that the distance estimated for the HESS source ($\sim$10 kpc) can be accepted as a lower bound on the distance to CXOU~J163802.6--471358 and is used as such for the remainder of this discussion. \\
\indent The photon index of $\Gamma= 1.1^{+0.7}_{-0.6}$ obtained from the joint {\em Chandra} and {\em XMM-Newton} fit is consistent with what is expected for a pulsar+PWN spectrum, where $\Gamma_{\text{PSR}}$ and $\Gamma_{\text{PWN}}$ usually lie in the range $1 \lesssim  \Gamma  \lesssim$ 2 \citep{kargaltsev2008a}. This value implies a somewhat harder spectrum than what is generally seen from an AGN \citep{mainieri2007} and the ACIS spectrum obtained for the point source is even harder ($\Gamma = 0.5^{+1.0}_{-0.9}$).\\
\indent Comparing the flux obtained from the {\em Chandra} observations, and the {\em XMM-Newton} observation done 5 years previously ($F^{Chandra}_{2-10\text{keV}}$=3.2$^{+0.9}_{-1.0}\times10^{-13}$ ergs cm$^{-2}$ s$^{-1}$ and $F^{XMM}_{2-10\text{keV}}$=2.6$^{+0.3}_{-0.8}\times10^{-13}$ ergs cm$^{-2}$ s$^{-1}$, respectively), no long timescale variability is seen. Since an AGN is expected to exhibit the strongest long timescale variability in the soft X-ray regime (< 12 keV) \citep{chitnis2009}, this result decreases the likelihood of the source being an AGN, unless we just happened to catch the source when it was at comparable flux levels. A LMXB is also expected to be strongly variable which makes this hypothesis less likely as well. The absence of X-ray variability on long timescale supports the PWN hypothesis, where usually only short time-scale variations at the pulse period are expected. \\
\indent We attempted a pulsation search with {\em XMM-Newton}, making use of the timing resolution of the PN instrument ($\sim$73 ms), unfortunately the poor photon statistics combined with the high background during the observation did not allow us to place any constraints on a possible pulsar period. \\
\indent Since the flux values from the two observation epochs ({\em Chandra} and {\em XMM-Newton}) are consistent, we used the unabsorbed flux obtained from the joint fit together with the distance given above in order to calculate a luminosity of the full source (point source + tail). This gives a luminosity in the 2-10 keV energy range of $\sim 4.8\times10^{33}d_{10}^2$ergs s$^{-1}$ which is in good agreement with the values measured for several pulsars (Table 1 and 2 in \citealt{kargaltsev2008a}).\\
\indent A radio counterpart to the X-ray source CXOU~J163802.6--471358 was found from MGPS-2 data, and revealed a radio trail aligned with the X-ray trail (Figure \ref{fig:radio}), which indicates that there is a connection between the two. The synchrotron origin of the emission suggested by the spectrum of the source is further supported by the fact that the radio emission stretches further to the North than does the X-ray emission. This could be the result of synchrotron aging effects, meaning that the low energy emitting electrons live longer than the high energy emitting electrons and therefore are able to travel further away from the origin of the emission (the point source). If this is indeed a radio counterpart to the X-ray source, the likelihood of the source being a LMXB is small, since only a very rare type of microquasar, such as 1E 1740.7--2942 or GRS 1758--258 \citep{mirabel1993}, with persistent radio jets would be likely to be found with extended radio emission. However, even in this LMXB scenario, the properties of CXOU~J163802.6--471358 are not a good match since the persistent radio jets are double-sided and the X-ray emission from most LMXBs is highly variable on a wide range of timescales, making it extremely unlikely that the {\em Chandra} and {\em XMM-Newton} observations, separated by six years, would have the same or nearly the same flux. Some exceptions do exist where the LMXBs are in a quiescent phase, only accreting at a low level and therefore exhibiting long periods of relatively stable luminosity, which typically lies in the range $L_X=10^{30}-10^{33}$~ergs s$^{-1}$ \citep{garcia2001}. The luminosity of $L_X\sim 4.8\times10^{33}d_{10}^2$ergs s$^{-1}$ for CXOU~J163802.6--471358 indicates that if this was a LMXB in its quiescent phase, it would be at the high end of the expected luminosity range for such sources, where significant variability is known to occur (see \citealt{cackett2011} and \citealt{bradley2007} for examples).   \\
\indent An offset between the radio and X-ray peak of $\sim40\arcsec$ is observed, and given the absolute astrometric offset of $< 2\arcsec$ for the radio survey \citep{murphy2007} and the $0.6\arcsec$ positional uncertainty for the {\em Chandra} source, the offset is real, assuming that the radio emission is in fact a counterpart to the X-ray source. Such an offset is seen for some PWNe driven by pulsars with high kick velocity (PSR~J1509--5850 in \citealt{kargaltsev2008b}, IGR J11014--6103 in \citealt{pavan2013}). Such an offset is not expected for an AGN where the nucleus should be a static emitting source with the radio peak coincident with the X-ray peak.  \\
\indent The nearest IR source is located 1.7$\arcsec$ from the X-ray source (Figure \ref{fig:IR}), and taking the positional error of $0.6\arcsec$ (white circle in Figure \ref{fig:IR}) on the {\em Chandra} source position into account, the IR source can be excluded as a counterpart to the X-ray source. This is consistent with the PWN interpretation, where the absence of an IR counterpart is expected. However, it is not an extremely strong constraint since any counterpart would be hidden by the large extinction towards the Norma region. \\
\indent We exclude the possibility of the source being a LMXB based on the above, and we argue that an AGN nature of the source is unlikely. On the other hand, no results of this analysis challenges the hypothesis of the point source being a high velocity pulsar driving a ram pressured PWN, which makes us believe that this is indeed a previously unknown bowshock-tail PWN. \\
\indent Based on the bowshock-tail interpretation of the source, we can use the empirical relation log~$L_{x,(2-10~\text{keV})}=1.34~$log~$\dot{E}-15.34$ given in \citet{possenti2002} to calculate the spin-down luminosity of the pulsar from the total unabsorbed source luminosity given above, which results in an estimate of the spin-down luminosity of $\dot{E}=4\times10^{36}$ ergs s$^{-1}$. The estimate is highly uncertain due to the large scatter in the above relation, and could be off by a factor of 10. However, the value seems reasonable for a bowshock tail PWN (see Table 1 in \citealt{kargaltsev2008a}), and is also close to the value found through radio timing observations of the pulsar PSR B1757--24 ($\dot{E}\sim2.6\times10^{36}$ergs s$^{-1}$; \citealt{kaspi2001}), which exhibits a very similar X-ray structure to what is observed for this source. We do not present other estimated values for the pulsar spin-parameters, since a careful examination of the relevant calculations reveals that the results vary a great deal depending on the values assumed. \\
\indent Using the 10\,kpc value as a lower bound on the distance to the source, and using the angular length of the X-ray tail of $\sim 40\arcsec$, the projected length of the tail is found to be $l_{\perp}\sim1.9d_{10}$ pc. The true length of the X-ray tail could be significantly longer, since we did not detect any significant change in tail length with energy, or significant softening of the spectrum with increasing distance from the pulsar candidate, which would otherwise be expected for a synchrotron emitting source. This might be due to the poor photon statistics that would lead to detection of only the brightest part of the tail where no effects of synchrotron aging would be detectable.\\
\indent Accepting the pulsar interpretation, makes it even more interesting to determine whether the jet-like feature perpendicular to the trail of the source (marked by the dashed ellipse in Figure \ref{fig:specreg}) is real or not. Associations could be drawn towards other pulsars showing evidence of a jet protruding almost perpendicular to the velocity vector of the pulsar (i.e. IGR J11014--6103 in \citealt{pavan2013} and B2224+65 in \citealt{johnson2010}). \\
\indent In order to prove or disprove the claim that the point source is a pulsar and that the extended emission is a bowshock tail created by ram pressure, we need longer exposure with {\em Chandra} and better resolution radio data. This would also provide us with the means of resolving the diffuse jet-like feature. A search for pulsations from the point source in radio or in X-ray using {\em XMM-Newton}, or {\em Chandra} in timing mode, would be very beneficial, and would provide the means to definitively determine the nature of the source. Making use of the increased resolution at hard X-ray energies provided by the {\em Nuclear Spectroscopic Telescope Array (NuSTAR)} we could possibly provide better constraints on the spectrum of this hard source.

\begin{acknowledgements}

The Dark Cosmology Centre is funded by the DNRF. JAT acknowledges partial support from NASA through {\em Chandra} Award G01-12068A issued by the {\em Chandra} X-ray Observatory Center, which is operated by the Smithsonian Astrophysical Observatory for and on behalf of NASA under contract NAS8-03060. This work is based in part on observations made with the {\em Spitzer} Space Telescope, which is operated by the Jet Propulsion Laboratory, California Institute of Technology under a contract with NASA.

\end{acknowledgements}

\bibliography{Normapaper_rev_arxiv}

\begin{thebibliography}{27}
\expandafter\ifx\csname natexlab\endcsname\relax\def\natexlab#1{#1}\fi

\bibitem[{{Benjamin} {et~al.}(2003){Benjamin}, {Churchwell}, {Babler}, {Bania},
  {Clemens}, {Cohen}, {Dickey}, {Indebetouw}, {Jackson}, {Kobulnicky},
  {Lazarian}, {Marston}, {Mathis}, {Meade}, {Seager}, {Stolovy}, {Watson},
  {Whitney}, {Wolff}, \& {Wolfire}}]{benjamin2003}
{Benjamin}, R.~A., {Churchwell}, E., {Babler}, B.~L., {et~al.} 2003, \pasp,
  115, 953

\bibitem[{{Bradley} {et~al.}(2007){Bradley}, {Hynes}, {Kong}, {Haswell},
  {Casares}, \& {Gallo}}]{bradley2007}
{Bradley}, C.~K., {Hynes}, R.~I., {Kong}, A.~K.~H., {et~al.} 2007, \apj, 667,
  427

\bibitem[{{Cackett} {et~al.}(2011){Cackett}, {Fridriksson}, {Homan}, {Miller},
  \& {Wijnands}}]{cackett2011}
{Cackett}, E.~M., {Fridriksson}, J.~K., {Homan}, J., {Miller}, J.~M., \&
  {Wijnands}, R. 2011, \mnras, 414, 3006

\bibitem[{{Cash}(1979)}]{cash1979}
{Cash}, W. 1979, \apj, 228, 939

\bibitem[{{Chitnis} {et~al.}(2009){Chitnis}, {Pendharkar}, {Bose}, {Agrawal},
  {Rao}, \& {Misra}}]{chitnis2009}
{Chitnis}, V.~R., {Pendharkar}, J.~K., {Bose}, D., {et~al.} 2009, \apj, 698,
  1207

\bibitem[{{Dame} {et~al.}(2001){Dame}, {Hartmann}, \& {Thaddeus}}]{dame2001}
{Dame}, T.~M., {Hartmann}, D., \& {Thaddeus}, P. 2001, \apj, 547, 792

\bibitem[{{Ferrand} \& {Safi-Harb}(2012)}]{ferrand2012}
{Ferrand}, G., \& {Safi-Harb}, S. 2012, Advances in Space Research, 49, 1313

\bibitem[{{Gaensler} \& {Slane}(2006)}]{gaensler2006}
{Gaensler}, B.~M., \& {Slane}, P.~O. 2006, \araa, 44, 17

\bibitem[{{Gaensler} {et~al.}(2004){Gaensler}, {van der Swaluw}, {Camilo},
  {Kaspi}, {Baganoff}, {Yusef-Zadeh}, \& {Manchester}}]{gaensler2004}
{Gaensler}, B.~M., {van der Swaluw}, E., {Camilo}, F., {et~al.} 2004, \apj,
  616, 383

\bibitem[{{Garcia} {et~al.}(2001){Garcia}, {McClintock}, {Narayan}, {Callanan},
  {Barret}, \& {Murray}}]{garcia2001}
{Garcia}, M.~R., {McClintock}, J.~E., {Narayan}, R., {et~al.} 2001, \apjl, 553,
  L47

\bibitem[{{Green}(2009)}]{green2009}
{Green}, D.~A. 2009, Bulletin of the Astronomical Society of India, 37, 45

\bibitem[{{Johnson} \& {Wang}(2010)}]{johnson2010}
{Johnson}, S.~P., \& {Wang}, Q.~D. 2010, \mnras, 408, 1216

\bibitem[{{Kargaltsev} {et~al.}(2008){Kargaltsev}, {Misanovic}, {Pavlov},
  {Wong}, \& {Garmire}}]{kargaltsev2008b}
{Kargaltsev}, O., {Misanovic}, Z., {Pavlov}, G.~G., {Wong}, J.~A., \&
  {Garmire}, G.~P. 2008, \apj, 684, 542

\bibitem[{{Kargaltsev} \& {Pavlov}(2008)}]{kargaltsev2008a}
{Kargaltsev}, O., \& {Pavlov}, G.~G. 2008, in American Institute of Physics
  Conference Series, Vol. 983, 40 Years of Pulsars: Millisecond Pulsars,
  Magnetars and More, ed. C.~{Bassa}, Z.~{Wang}, A.~{Cumming}, \& V.~M.
  {Kaspi}, 171--185

\bibitem[{{Kaspi} {et~al.}(2001){Kaspi}, {Gotthelf}, {Gaensler}, \&
  {Lyutikov}}]{kaspi2001}
{Kaspi}, V.~M., {Gotthelf}, E.~V., {Gaensler}, B.~M., \& {Lyutikov}, M. 2001,
  \apjl, 562, L163

\bibitem[{{Kirk} {et~al.}(2009){Kirk}, {Lyubarsky}, \& {Petri}}]{kirk2009}
{Kirk}, J.~G., {Lyubarsky}, Y., \& {Petri}, J. 2009, in Astrophysics and Space
  Science Library, Vol. 357, Astrophysics and Space Science Library, ed.
  W.~{Becker}, 421

\bibitem[{{Lemiere} {et~al.}(2009){Lemiere}, {Slane}, {Gaensler}, \&
  {Murray}}]{Lemiere2009}
{Lemiere}, A., {Slane}, P., {Gaensler}, B.~M., \& {Murray}, S. 2009, \apj, 706,
  1269

\bibitem[{{Mainieri} {et~al.}(2007){Mainieri}, {Hasinger}, {Cappelluti},
  {Brusa}, {Brunner}, {Civano}, {Comastri}, {Elvis}, {Finoguenov}, {Fiore},
  {Gilli}, {Lehmann}, {Silverman}, {Tasca}, {Vignali}, {Zamorani},
  {Schinnerer}, {Impey}, {Trump}, {Lilly}, {Maier}, {Griffiths}, {Miyaji},
  {Capak}, {Koekemoer}, {Scoville}, {Shopbell}, \& {Taniguchi}}]{mainieri2007}
{Mainieri}, V., {Hasinger}, G., {Cappelluti}, N., {et~al.} 2007, \apjs, 172,
  368

\bibitem[{{Minniti} {et~al.}(2010){Minniti}, {Lucas}, {Emerson}, {Saito},
  {Hempel}, {Pietrukowicz}, {Ahumada}, {Alonso}, {Alonso-Garcia}, {Arias},
  {Bandyopadhyay}, {Barb{\'a}}, {Barbuy}, {Bedin}, {Bica}, {Borissova},
  {Bronfman}, {Carraro}, {Catelan}, {Clari{\'a}}, {Cross}, {de Grijs},
  {D{\'e}k{\'a}ny}, {Drew}, {Fari{\~n}a}, {Feinstein}, {Laj{\'u}s}, {Gamen},
  {Geisler}, {Gieren}, {Goldman}, {Gonzalez}, {Gunthardt}, {Gurovich},
  {Hambly}, {Irwin}, {Ivanov}, {Jord{\'a}n}, {Kerins}, {Kinemuchi}, {Kurtev},
  {L{\'o}pez-Corredoira}, {Maccarone}, {Masetti}, {Merlo}, {Messineo},
  {Mirabel}, {Monaco}, {Morelli}, {Padilla}, {Palma}, {Parisi}, {Pignata},
  {Rejkuba}, {Roman-Lopes}, {Sale}, {Schreiber}, {Schr{\"o}der}, {Smith}, {},
  {Soto}, {Tamura}, {Tappert}, {Thompson}, {Toledo}, {Zoccali}, \&
  {Pietrzynski}}]{minniti2010}
{Minniti}, D., {Lucas}, P.~W., {Emerson}, J.~P., {et~al.} 2010, New A, 15, 433

\bibitem[{{Mirabel} {et~al.}(1993){Mirabel}, {Rodriguez}, {Cordier}, {Paul}, \&
  {Lebrun}}]{mirabel1993}
{Mirabel}, I.~F., {Rodriguez}, L.~F., {Cordier}, B., {Paul}, J., \& {Lebrun},
  F. 1993, \aaps, 97, 193

\bibitem[{{Murphy} {et~al.}(2007){Murphy}, {Mauch}, {Green}, {Hunstead},
  {Piestrzynska}, {Kels}, \& {Sztajer}}]{murphy2007}
{Murphy}, T., {Mauch}, T., {Green}, A., {et~al.} 2007, \mnras, 382, 382

\bibitem[{{Pavan} {et~al.}(2013){Pavan}, {Bordas}, {Puehlhofer}, {Filipovic},
  {De Horta}, {O'Brien}, {Balbo}, {Walter}, {Bozzo}, {Ferrigno}, {Crawford}, \&
  {Stella}}]{pavan2013}
{Pavan}, L., {Bordas}, P., {Puehlhofer}, G., {et~al.} 2013, ArXiv e-prints

\bibitem[{{Possenti} {et~al.}(2002){Possenti}, {Cerutti}, {Colpi}, \&
  {Mereghetti}}]{possenti2002}
{Possenti}, A., {Cerutti}, R., {Colpi}, M., \& {Mereghetti}, S. 2002, \aap,
  387, 993

\bibitem[{{Saito} {et~al.}(2012){Saito}, {Hempel}, {Minniti}, {Lucas},
  {Rejkuba}, {Toledo}, {Gonzalez}, {Alonso-Garc{\'{\i}}a}, {Irwin},
  {Gonzalez-Solares}, {Hodgkin}, {Lewis}, {Cross}, {Ivanov}, {Kerins},
  {Emerson}, {Soto}, {Am{\^o}res}, {Gurovich}, {D{\'e}k{\'a}ny}, {Angeloni},
  {Beamin}, {Catelan}, {Padilla}, {Zoccali}, {Pietrukowicz}, {Moni Bidin},
  {Mauro}, {Geisler}, {Folkes}, {Sale}, {Borissova}, {Kurtev}, {Ahumada},
  {Alonso}, {Adamson}, {Arias}, {Bandyopadhyay}, {Barb{\'a}}, {Barbuy},
  {Baume}, {Bedin}, {Bellini}, {Benjamin}, {Bica}, {Bonatto}, {Bronfman},
  {Carraro}, {Chen{\`e}}, {Clari{\'a}}, {Clarke}, {Contreras}, {Corvill{\'o}n},
  {de Grijs}, {Dias}, {Drew}, {Fari{\~n}a}, {Feinstein},
  {Fern{\'a}ndez-Laj{\'u}s}, {Gamen}, {Gieren}, {Goldman},
  {Gonz{\'a}lez-Fern{\'a}ndez}, {Grand}, {Gunthardt}, {Hambly}, {Hanson},
  {He{\l}miniak}, {Hoare}, {Huckvale}, {Jord{\'a}n}, {Kinemuchi}, {Longmore},
  {L{\'o}pez-Corredoira}, {Maccarone}, {Majaess}, {Mart{\'{\i}}n}, {Masetti},
  {Mennickent}, {Mirabel}, {Monaco}, {Morelli}, {Motta}, {Palma}, {Parisi},
  {Parker}, {Pe{\~n}aloza}, {Pietrzy{\'n}ski}, {Pignata}, {Popescu}, {Read},
  {Rojas}, {Roman-Lopes}, {Ruiz}, {Saviane}, {Schreiber}, {Schr{\"o}der},
  {Sharma}, {Smith}, {Sodr{\'e}}, {Stead}, {Stephens}, {Tamura}, {Tappert},
  {Thompson}, {Valenti}, {Vanzi}, {Walton}, {Weidmann}, \&
  {Zijlstra}}]{saito2012}
{Saito}, R.~K., {Hempel}, M., {Minniti}, D., {et~al.} 2012, \aap, 537, A107

\bibitem[{{Schlafly} \& {Finkbeiner}(2011)}]{schlafly2011}
{Schlafly}, E.~F., \& {Finkbeiner}, D.~P. 2011, \apj, 737, 103

\bibitem[{{Watson}(2011)}]{watson2011}
{Watson}, D. 2011, \aap, 533, A16

\bibitem[{{Wilms} {et~al.}(2000){Wilms}, {Allen}, \& {McCray}}]{wilms2000}
{Wilms}, J., {Allen}, A., \& {McCray}, R. 2000, \apj, 542, 914

\end{thebibliography}

\end{document}